\documentclass[prb, reprint, amsfonts, amssymb, amsmath, superscriptaddress]{revtex4-2}
\usepackage[utf8]{inputenc}
\usepackage{braket}
\usepackage{graphicx}
\usepackage{xcolor}
\usepackage{hyperref}
\hypersetup{colorlinks, 
	linkcolor={blue!75!black!80!yellow},
	citecolor={blue!75!black!80!yellow}, 
	urlcolor={blue!75!black!80!yellow}
}

\let\hatOrig\hat
\renewcommand{\vec}[1]{{\boldsymbol{\mathbf{#1}}}}
\renewcommand{\hat}[1]{{\boldsymbol{\mathbf{\hatOrig{#1}}}}}
\renewcommand{\Im}{\operatorname{Im}}
\renewcommand{\Re}{\operatorname{Re}}

\newcommand{\sub}[1]{\ensuremath{_{\textrm{#1}}}}   \newcommand{\mesh}[3]{\ensuremath{#1 \times #2 \times #3}}

\newcommand{\affilMSE}{\affiliation{Department of Materials Science and Engineering, Rensselaer Polytechnic Institute, Troy, NY 12180, USA}}
\newcommand{\affilPhy}{\affiliation{Department of Physics, Applied Physics, and Astronomy, Rensselaer Polytechnic Institute, Troy, NY 12180, USA}}
\newcommand{\affilUWM}{\affiliation{Department of Materials Science and Engineering, University of Wisconsin-Madison, WI, 53706, USA}}
\newcommand{\affilUCSC}{\affiliation{Department of Physics, University of California, Santa Cruz, CA, 95064, USA}}
\newcommand{\affiljxu}{\affiliation{Department of Physics, Hefei University of Technology, Hefei, Anhui, China}}
\newcommand{\affilUCSChem}{\affiliation{Department of Chemistry and Biochemistry, University of California, Santa Cruz, CA, 95064, USA}}

\begin{document}
\title{Circular Dichroism of Crystals from First Principles}

\author{Christian Multunas}\affilPhy
\author{Andrew Grieder} \affilUWM
\author{Junqing Xu}\affiljxu\affilUCSChem
\author{Yuan Ping}\email{yping3@wisc.edu} \affilUWM\affilUCSC
\author{Ravishankar Sundararaman}\email{sundar@rpi.edu}\affilPhy\affilMSE
\date{\today}

\begin{abstract}
Chiral crystals show promise for spintronic technologies on account of their high spin selectivity, which has led to significant recent interest in quantitative characterization and first-principles prediction of their spin-optoelectronics properties.
Here, we outline a computational framework for efficient \emph{ab-initio} calculations of circular dichroism (CD) in crystalline materials. We leverage direct calculations of orbital angular momentum and quadrupole matrix element calculations in density-functional theory (DFT) and Wannier interpolation to calculate CD in complex materials, removing the need for band convergence and accelerating Brillouin-zone convergence compared to prior approaches.
We find strong agreement with measured CD signals in molecules and crystals ranging in complexity from small bulk unit cells to 2D hybrid perovskites, and show the importance of the quadrupole contribution to the anisotropic CD in crystals.
Spin-orbit coupling affects the CD of crystals with heavier atoms, as expected, but this is primarily due to changes in the electronic energies, rather than due to direct contributions from the spin matrix elements.
We showcase the capability to predict CD for complex structures on a 2D hybrid perovskite, finding strong orientation dependence and identifying the eigen-directions of the unit cell with the strongest CD.
We additionally decompose CD into separate contributions from inorganic, organic, and mixed organic-inorganic transitions, finding the chiral molecules to dominate the CD, with the inorganic lattice contributing at higher frequencies in specific directions.
This unprecedented level of detail in CD predictions in crystals will facilitate experimental development of complex chiral crystals for spin selectivity.

\end{abstract}


\maketitle

\section{Introduction}  \label{sec:intro}

Chiral materials have recently received significant attention due to the unique electronic structure arising from the combination of chiral symmetry-breaking and spin-orbit coupling.
Most notably, chiral non-magnetic materials generate a spin-polarized current when flowing a charge current, referred to as Chirality-Induced Spin Selectivity (CISS)~\cite{Naaman2012}.
Chiral crystals show particular promise for spintronic devices because of significantly greater CISS magnitude, compared to molecules~\cite{Lu2019, Lu2020, Lu2021}.
In addition, chiral materials can emit strongly circularly-polarized light for spin-polarized light-emitting diodes (spin-LED)~\cite{Young2021}.

This renewed interest in chirality of materials necessitates wide-ranging techniques to benchmark their structure, electronic and optical properties, computationally and experimentally.
Chiral crystals could potentially demonstrate connections between chiral optical properties, which are typically easier to measure, and their spin transport properties, facilitating faster identification of candidate chiral materials for spintronics.
In particular, the standard experimental test of chirality in materials is the difference in optical response between left- and right-circularly polarized (LC and RC) light.
Specifically, circular dichroism (CD) is the differential absorbance spectrum, $\Delta A(\omega) = A\sub{LC}(\omega) - A\sub{RC}(\omega)$, which is equal and opposite for a pair of enantiomers (chiral structures connected by a reflection).
CD spectroscopy is well-established in organic chemistry and biomaterials due to the prevalence of chiral asymmetry in molecules, and is routinely applied to identify enantiomers and classify secondary structures~\cite{Kelly2005, Greenfield1969}.

CD measurements of molecules in gas-phase or solution are typically easier to perform than for chiral crystals.
For crystals, the typically stronger attenuation and the presence of crystalline anisotropy require careful choice of sample dimensions and orientation during measurement, especially alignment with optical axes~\cite{Castiglioni2009}. 
With increasingly complex chiral crystal structures, first-principles prediction of CD is necessary to aid experimental identification and characterization of chiral crystals.
Most prior first-principles calculations of CD focus on molecules~\cite{Rogers2004}, where the signal is averaged over all molecule orientations.
These frameworks ignore crystal anisotropy~\cite{Ganyushin2008}, and additional contributions from electric quadrupole matrix elements~\cite{Govorov2010, Warnke2012}, that are important for the orientation dependence.
Recent developments have accounted for these effects in first-principles calculations of chiral periodic systems, finding good agreement with orientation-dependent CD measurements for chiral crystals such as trigonal tellurium~\cite{Wang2023}.

A pending challenge is extending such techniques to complex crystal structures, and enable rapid evaluation of several chiral material candidates.
The current state-of-the-art requires summation over several empty bands to compute orbital angular momentum and electric quadrupole matrix elements, and simultaneously dense Brillouin-zone sampling, to converge the predicted CD spectrum~\cite{Wang2023}.
This incurs a high computational cost for predicting CD and optical activity in materials with large unit cells, such as for hybrid organic-inorganic perovskites (HOIPs).
These complex systems have drawn significant recent attention for exceptional opto-spintronic properties~\cite{Lu2019,Lu2020,Lu2021,Saparov2016,Kagan1999}, and routinely involve 100-200 atoms and over 1000 valence electrons per unit cell, necessitating further advances for efficient first-principles CD predictions.

Here, we develop an \emph{ab initio} approach for efficiently predicting orientation-dependent CD in crystals.
By directly computing angular momentum and quadrupole matrix elements without relying on a sum over states, we eliminate the need for convergence with respect to large number of empty states.
We also simultaneously leverage Wannier interpolation of these matrix elements to enable rapid convergence of Brillouin zone integration.
Together, this allows us to inexpensively compute CD in chiral crystals, varying in complexity from elemental Te to 2D hybrid perovskites, and showing excellent agreement with experiment across this range, ensuring the wide-scale reliability of this approach.
Using this framework, we showcase the importance of the quadrupole matrix element contribution for orientation-dependent CD in anisotropic crystals.
We also determine the impact of spin-orbit coupling (SOC) on CD of materials with heavy atoms, and show that spin contributes primarily by modifying the band structure, rather than by directly through spin magnetic dipole matrix elements.
 \section{Theory}

In the following sections, we extensively detail the complete theoretical formulation and computational implementation of CD calculations in crystals.
We first establish the necessary optical matrix elements, including magnetic dipole (orbital and spin) and electric quadrupole contributions, and using them, formulate CD as a rank-2 tensor with respect to propagation direction.
We then discuss calculation of the necessary matrix elements using two complementary approaches, direct calculation in density-functional theory (DFT) using derivatives with respect to the Bloch wave-vector $\vec{k}$, and using Wannier interpolation to accelerate Brillouin zone integration.

\subsection{Higher-order optical transition matrix elements}

In the isotropic regime of molecules in gas phase or solution, the standard approach for predicting optical cavity shows that CD $\propto \Im[\vec{\mu}_{0n} \cdot \vec{m}_{0n}]$, where $\vec{\mu}_{0n}$ and $\vec{m}_{0n}$ are respectively the electric dipole and magnetic dipole transition matrix elements (between ground state 0 and excited state $n$)~\cite{Govorov2010}.
When examining CD response in chiral crystals, it becomes important to consider anisotropy of optical response along different crystallographic orientations.
Chirality results from a lack of inversion and mirror symmetries, generally leading to crystals with low symmetries that are likely to exhibit a strongly direction-dependent CD spectrum.
For completeness, we outline below the derivation of the orientation-dependent CD and the matrix elements that it involves.

We start with the Pauli Hamiltonian for electrons in an external electromagnetic field,
\begin{equation}
\hat{H} = \frac{(\vec{p} + e\vec{A})^2 + e\hbar\vec{\sigma} \cdot (\nabla \times \vec{A})} {2m}
\end{equation}
with electron momentum $\vec{p}$ and vector potential $\vec{A} = \vec{A}_0e^{i(\vec{q}\cdot\vec{r}-\omega t)}$ of an electromagnetic wave propagating with wave vector $\vec{q}$.
We can extract the perturbation Hamiltonian at linear order in $A_0$, which will only exhibit transition matrix elements between electronic states with $\vec{k}$ differing by photon wave vector $\vec{q}$.
Further, since the light wavelength is much longer than the crystal, we expand in powers of $q$, but retain terms up to $O(q)$ instead of retaining only $q$-independent terms as in the usual dipole approximation.
This leads to the transition matrix elements between bands $n$ and $n'$ near wavevector $\vec{k}$ at $O(A_0)$ given by
\begin{multline}
\bra{\vec{k}+\vec{q},n'} \hat{H}' \ket{\vec{k},n} \\
= \frac{e}{m}
\left(
    \vec{A}_0 \cdot \vec{p}^\vec{k}_{n'n}
    + i q_\mu A_{0\nu} X^\vec{k}_{\mu\nu,n'n}
    + \vec{B}_0 \cdot \vec{S}^\vec{k}_{n'n}
\right) + O(q^2),
\end{multline}
where repeated indices are summed per the Einstein convention and magnetic field $\vec{B}_0 \equiv \nabla\times\vec{A}_0$.
Above, $\vec{p}$ and $\vec{S}$ are the momentum and spin matrix elements, while the tensor
\begin{equation}
X^\vec{k}_{\mu\nu,n'n} \equiv \frac{
\bra{i \partial_{k_\mu} u_{\vec{k}n'} } p_\nu 
\ket{u_{\vec{k}n}} + h.c.}{2},
\label{eq:X}
\end{equation}
where $u_{\vec{k}n}(\vec{r})$ is the Bloch function for band $n$ at wavevector $\vec{k}$,
includes both the orbital angular momentum
\begin{equation}
    L_\rho = \epsilon_{\rho\mu\nu} X_{\mu\nu}
\label{eq:L}
\end{equation}
as the antisymmetric part (where $\epsilon_{\rho\mu\nu}$ is the Levi-Civita tensor), and the electric quadrupole 
\begin{equation}
    Q_{\mu\nu} = X_{\mu\nu} + X_{\nu\mu} - \frac{2}{3} \delta_{\mu\nu} X_{\rho\rho}
\label{eq:Q}
\end{equation}
as the traceless symmetric part.
Substituting $X$ in terms of $L$ and $Q$, we can write the overall transition matrix element as
\begin{equation}
\bra{\vec{k}+\vec{q},n'} \hat{H}' \ket{\vec{k},n} = 
\frac{eA_{0\nu}}{m} \left( p_{\nu,n'n}^\vec{k} + \frac{iq_\mu}{2} Y_{\mu\nu,n'n}^\vec{k} \right),
\end{equation}
where 
\begin{equation}
Y_{\mu\nu,n'n}^\vec{k} \equiv Q_{\mu\nu,n'n}^\vec{k} + \epsilon_{\rho\mu\nu} ( L_{\rho,n'n}^\vec{k} + 2 S_{\rho,n'n}^\vec{k} )
\label{eq:Y}
\end{equation}
collects together all matrix elements at $O(q)$, \textit{i.e.}, beyond the dipole approximation.

\subsection{Differential absorbance tensor}

Applying Fermi's Golden Rule to the above transition matrix elements, we can calculate the imaginary part of the dielectric function and expand it in powers of $q$ as.
\begin{equation}
\Im \epsilon_{\alpha\beta}=
    \Im \epsilon_{\alpha\beta}^0 + iq_\mu \Im \epsilon_{\alpha\beta\mu}^1 + O(q^2).
\label{eq:ImEpsSeries}
\end{equation}
This yields the standard dielectric function with only dipole contributions at $O(q^0)$,
\begin{multline}
    \Im \epsilon_{\alpha\beta}^0 \equiv
    \frac{4\pi^2e^2}{m_e^2\omega^2} \int_{BZ} \frac{g_sd\vec{k}}{(2\pi)^3} 
    \sum_{n',n} \delta(\varepsilon_{\vec{k}n'}-\varepsilon_{\vec{k}n}-\hbar\omega) \\
    \times (f_{\vec{k}n} - f_{\vec{k}n'})
    (p_{\alpha,n'n}^{\vec{k}*}p_{\beta,n'n}^{\vec{k}}),
\label{eq:ImEps0}
\end{multline}
where $\varepsilon_{\vec{k}n}$ and $f_{\vec{k}n}$ are the single-particle electron energies and Fermi occupation factors for band $n$ and wavevector $\vec{k}$, and $g_s$ is the spin-degeneracy factor (1 for spinorial, and 2 for non-spinorial unpolarized calculations)~\cite{Brown2016}.
Meanwhile, the $O(q^1)$ coefficient in Eq.~\ref{eq:ImEpsSeries} given by
\begin{multline}
\Im \epsilon_{\alpha\beta\mu}^1 \equiv
    \frac{4\pi^2e^2}{m_e^2\omega^2} \int_{BZ} \frac{g_sd\vec{k}}{(2\pi)^3} 
    \sum_{n',n} \delta(\varepsilon_{\vec{k}n'}-\varepsilon_{\vec{k}n}-\hbar\omega)\\
    \times (f_{\vec{k}n} - f_{\vec{k}n'})
    \frac{p_{\alpha,n'n}^{\vec{k}*}Y_{\mu\beta,n'n}^{\vec{k}} - Y_{\mu\alpha,n'n}^{\vec{k}*}p_{\beta,n'n}^{\vec{k}}}{2}
\label{eq:ImEps1}
\end{multline}
encapsulates the CD contributions, with magnetic dipole and electric quadrupole contributions within $Y_{\mu\alpha,n'n}^{\vec{k}}$.

To compute the effective dielectric function for circularly polarized light, we need to contract with the complex vector potential amplitudes 
$\vec{A}_0^{LC}$ and $\vec{A}_0^{RC}$ for left and right-circularly polarized light.
For CD, the standard sign convention used to define LC and RC is $\vec{A}_0^{LC} = (\hat{x} + i\hat{y})/\sqrt{2}$ and $\vec{A}_0^{RC} = (\hat{x} - i\hat{y})/\sqrt{2}$ for a wave propagating along $\hat{q} = \hat{z}$.
Rotating these polarizations for an arbitrary propagation direction, we can show that $A_{0\alpha}^{LC*}A_{0\beta}^{LC} - A_{0\alpha}^{RC*}A_{0\beta}^{RC} = i \epsilon_{\alpha\beta\gamma} \hat{q}_\gamma$, leading to
\begin{equation}
\Im \epsilon^{LC} (\vec{q}) - \Im \epsilon^{RC} (\vec{q}) =
    \hat{q}_\mu \hat{q}_\nu (-\epsilon_{\alpha\beta\nu} q \Im \epsilon_{\alpha\beta\mu}^1).
\end{equation}
Above, the $\Im \epsilon_{\alpha\beta}^0$ term drops out because it is symmetric in $\alpha \leftrightarrow \beta$ and contracted against the antisymmetric Levi-Civita tensor.

Next, we need to convert dielectric functions to absorbance for predicting CD.
The complex dielectric function leads to a complex wavevector $q = (\omega/c)\sqrt{\epsilon(\omega)}$, which causes the intensity to fall off exponentially as $I = I_0 e^{-2\Im q l}$ with path length $l$.
Hence, the Naperian absorbance is $\alpha = 2\Im q = 2(\omega/c) \Im\sqrt{\epsilon}$.
In the limit of low loss tangent ($|\Im\epsilon| \ll |\epsilon|$), we can simplify this to $\alpha \approx (\omega/(c\sqrt{\epsilon}) \Im\epsilon = (\omega^2/(c^2q)) \Im\epsilon$.

We can then convert the difference in $\Im \epsilon$ to a differential absorbance,
\begin{flalign}
\Delta\alpha &\equiv \alpha^{LC}(\vec{q}) - \alpha^{RC}(\vec{q}) \nonumber\\
&= \hat{q}_\mu \hat{q}_\nu \operatorname{Sym}_{\mu\nu} \left[ 
-\epsilon_{\alpha\beta\nu} \frac{\omega^2}{c^2}
\Im \epsilon_{\alpha\beta\mu}^1 \right],
\label{eq:CD_is_a_tensor}
\end{flalign}
where $\operatorname{Sym}_{\mu\nu} [F_{\mu\nu}] \equiv \frac{F_{\mu\nu} + F_{\nu\mu}}{2}$ for any tensor $F_{\mu\nu}$.
Note that the magnitude of $q$, which depends on $\Re\epsilon$, only cancels out in the low loss tangent limit and must be explicitly calculated using $\Re\epsilon$ obtained using the Kramers-Kronig relation applied to Eq.~\ref{eq:ImEps0} for high loss cases, which we do not consider here.
Equation~\ref{eq:CD_is_a_tensor} shows that circular dichroism $\Delta\alpha = \Delta\alpha_{\mu\nu} \hat{q}_\mu \hat{q}_\nu$ is a symmetric rank-2 tensor with respect to propagation direction $\hat{q}$.
Importantly, this is in contrast to the dielectric function $\epsilon = \epsilon_{\mu\nu} \hat{E}_\mu \hat{E}_\nu$, which is a rank-2 tensor for crystals with respect to the electric field direction $\hat{E}$.

Finally, substituting Eqs.~\ref{eq:Y} and \ref{eq:ImEps1} into Eq.~\ref{eq:CD_is_a_tensor}, we arrive at the final expression for the circular dichroism tensor,
\begin{multline}
\Delta\alpha_{\mu\nu} = \frac{4\pi^2e^2}{m_e^2\omega^2} \int_{BZ}
\frac{g_s d\vec{k}}{(2\pi)^3} \sum_{n',n}
\delta(\varepsilon_{\vec{k}n'}-\varepsilon_{\vec{k}n}-\hbar\omega)
\\
\times (f_{\vec{k}n} - f_{\vec{k}n'}) \Re\operatorname{Sym} \left[ 
\delta_{\mu\nu} p^{\vec{k}*}_{\rho,n'n} (L^{\vec{k}}_{\rho,n'n} + 2S^{\vec{k}}_{\rho,n'n})
\right. \\ \hspace{10em} \left.
- p^{\vec{k}*}_{\mu,n'n} (L^{\vec{k}}_{\nu,n'n} + 2S^{\vec{k}}_{\nu,n'n})
\right. \\ \left.
+ \epsilon_{\mu\nu\rho} p^{\vec{k}*}_{\sigma,n'n} Q^{\vec{k}}_{\rho\sigma,n'n}
\right].
\label{eq:CD}
\end{multline}
In general, all matrix element terms contribute for crystals.
However, for isotropic systems or, for molecules after computing an orientation average, the tensor reduces to the scalar $(1/3)\operatorname{Tr}(\Delta \alpha)$.
The term depending on the quadrupole matrix elements $Q$ is traceless and vanishes, while the isotropic average becomes proportional to $\vec{p}^\ast\cdot (\vec{L}+2\vec{S})$, \textit{i.e.}, electric dipole $\cdot$ magnetic dipole, as is well known for molecules~\cite{Govorov2010}.

\subsection{Direct calculation of $L,Q$ matrix elements}
\label{sec:DirectLQ}

In order to calculate CD using Eq.~\ref{eq:CD}, for each wavevector $\vec{k}$ and band pairs $n,n'$, we need the matrix elements of momentum $\vec{p}$, spin $\vec{S}$ (if accounting for SOC), angular momentum $\vec{L}$ and electric quadrupole $Q_{\mu\nu}$.
Computation of $\vec{p}$ and $\vec{S}$ is straightforward directly in density-functional theory, as well as through Wannier interpolation.
However, $\vec{L}$ and $Q_{\mu\nu}$, given by Eqs.~\ref{eq:L} and \ref{eq:Q}, require evaluation of Bloch function derivatives $\partial_\vec{k} u_{\vec{k}n}(\vec{r})$ within $X_{\mu\nu}$ in Eq.~\ref{eq:X}, which is non-trivial.

The most common approach to deal with these Bloch function derivatives is to insert an identity operator resolved as a sum over all states $\sum_m \ket{u_{\vec{k}m}} \bra{u_{\vec{k}m}}$ within Eq.~\ref{eq:X}.
The resulting expectations of the form $\braket{i\partial_{\vec{k}} u_{\vec{k}n'} | u_{\vec{k}m}}$ can then be written in terms of momentum matrix elements using $\vec{p} \equiv (-im/\hbar) [\vec{r}, \hat{H}]$, leading to
\begin{equation}
X^\vec{k}_{\mu\nu,n'n} = \frac{\hbar}{2im} \sum_{\varepsilon_{\vec{k}m} \neq \varepsilon_{\vec{k}n}}
    \frac{p_{\mu,n'm}^\vec{k} p_{\nu,mn}^\vec{k}}{\varepsilon_{\vec{k}m} -\varepsilon_{\vec{k}n}} + h.c.
\label{eq:XfromP}
\end{equation}
This expression has the advantage that it only relies on single-particle energies and momentum matrix elements readily available from any DFT code, and is the basis of most previous methods to calculate CD~\cite{Hidalgo2009, Wang2023}.
However, the sum over states introduced necessitates calculating a large number of empty bands, and converging with respect to this number.
This convergence will be slow and cell-size dependent for lower-dimensional systems in particular.
Additionally, for large unit cells such as for the hybrid perovskites, this requirement substantially increases the computational cost.

As an alternative, we directly implement the Bloch wavevector derivative in DFT as a finite difference derivative over $\vec{k}$.
We effectively compute the derivatives $\partial_\vec{k} u_{\vec{k}n}(\vec{r})$ using
\begin{equation}
    d\vec{k} \cdot \partial_\vec{k} u_{\vec{k}n}(\vec{r}) = 
    \sum_{n'} u_{\vec{k}+d\vec{k},n'}(\vec{r})
    U_{n'n} - u_{\vec{k}n}(\vec{r}),
\end{equation}
where where matrix $U$ aligns the phases for non-degenerate bands and unitary rotations within degenerate subspaces between the Bloch functions computed at $\vec{k}$ and $\vec{k}+d\vec{k}$.
This is necessary because the DFT calculation of Bloch states at different $\vec{k}$ are undetermined up to these phases / unitary rotations in general.
Specifically, to compute $U$, we first compute the overlap matrix $O_{n'n} = \braket{u_{\vec{k}+d\vec{k},n'} | u_{\vec{k},n}}$ and then calculate
\begin{equation}
U_{n'n} = 
\begin{cases}
(\bar{U}\bar{V}^\dagger)_{n'n}, & |\varepsilon_{n'} - \varepsilon_{n}| < d\varepsilon \\
0, & \mathrm{otherwise},
\end{cases}
\end{equation}
where the rotation within each degenerate subspace is given by
\begin{equation}
\bar{U}_{ab} = 
\begin{cases}
(\tilde{O}(\tilde{O}\tilde{O}^\dagger)^{-1/2})_{ab}, & |v_{a} - v_{b}| < dv \\
0, & \mathrm{otherwise}
\end{cases}.
\end{equation}
Above, $\tilde{O}$ is the submatrix within the $v$-degenerate subspace of $\bar{O}\bar{V}$, where $\bar{O}$ is the submatrix of $O$ within the $\epsilon$-degenerate subspace, while $\bar{V}$ and $v$ are the eigenvectors and eigenvalues of the momentum operator projected along the perturbation direction, $d\vec{k}\cdot\vec{p}$.
We use thresholds of $d\varepsilon = 10^{-6}~E_h$ and $dv = 10^{-4}~E_ha_0$ for identifying the degeneracies (where $E_h$ is the Hartree energy and $a_0$ is the Bohr radius).
Essentially, this scheme uses the momentum operator first to resolve the energy degeneracy and align the phases, and when the momenta are also degenerate, computes a unitary rotation from symmetric orthornormalization of the overlap $\tilde{O}$.
This is also effectively the Hamiltonian-gauge equivalent of the Wannier-gauge scheme for computing Berry curvature and related properties introduced in~\cite{Yates2007}.

For each $\vec{k}$, the above process is repeated with six $d\vec{k}$ along each (positive and negative) Cartesian direction in reciprocal space, each with magnitude $|d\vec{k}| = 10^{-4} a_0^{-1}$.
The $\partial_{\vec{k}} u_{\vec{k}n}$ computed from the positive and negative displacement for each direction are averaged together, resulting effectively in a central-difference derivative evaluation (with phase/rotation matching as outlined above).
After obtaining $\partial_{k_\mu} u_{\vec{k}n'}$, we compute its momentum $p_\nu$ matrix elements against all $u_{\vec{k}n}$ to get the $X^{\vec{k}}_{\mu\nu,n'n}$ defined in Eq.~\ref{eq:X}.
Note that the momentum operator is defined using the $[\vec{r},\hat{H}]$ commutator to properly account for non-local pseudopotential contributions~\cite{Brown2016, Sundararaman2017}.

\begin{figure}
\includegraphics[width=2.94in]{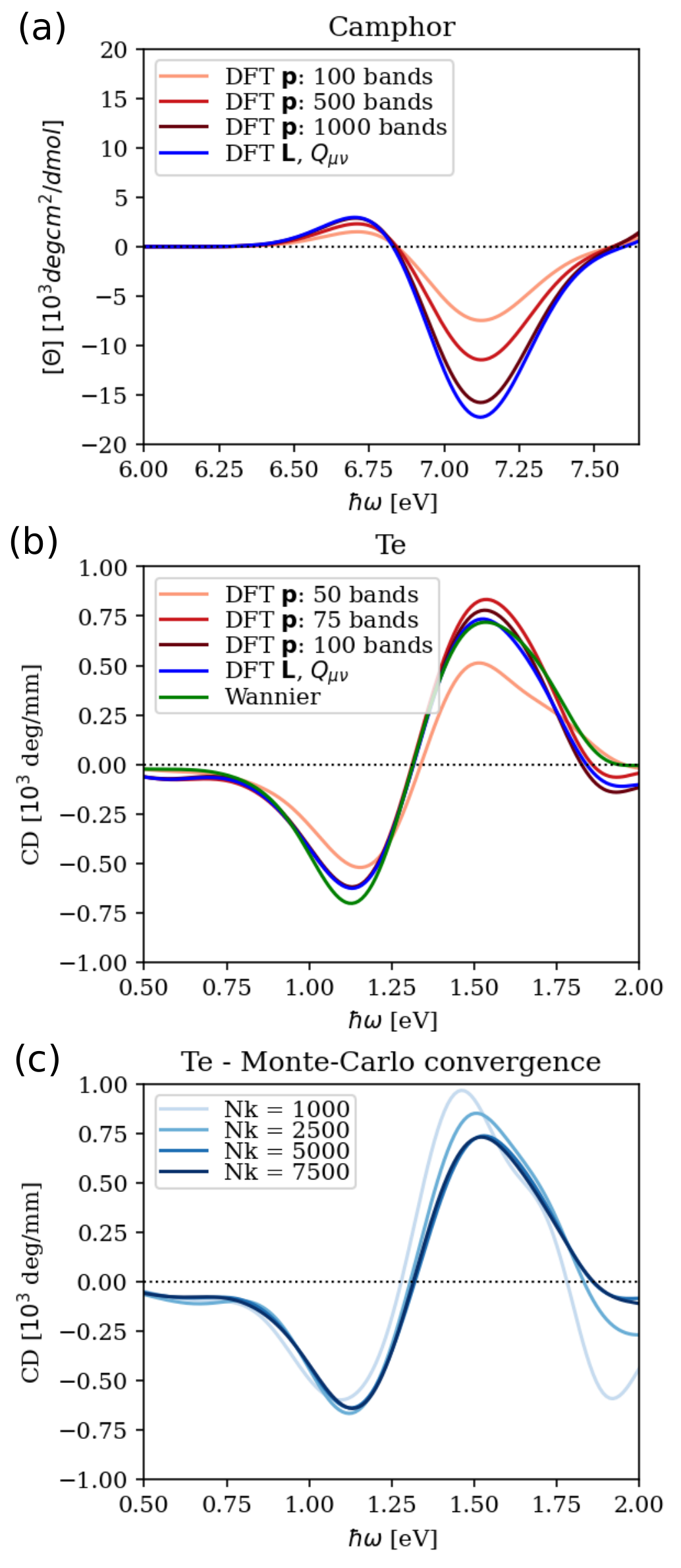}
\caption{Band convergence of CD computed purely from momentum $\vec{p}$ matrix elements using Eq.~\ref{eq:XfromP}, compared to the direct DFT evaluation of $\vec{L}$ and $Q_{\mu\nu}$ matrix elements without the additional sum over bands, for (a) camphor and (b) Te, which additionally shows the Wannier prediction for comparison.
More than 1000 bands are necessary to converge the result for the camphor molecule, while 100 bands achieve adequate convergence for the Te crystal.
(c) Convergence of direct DFT prediction of CD for Te with number of $k$-points for Brillouin zone integration.}
\label{fig:conv}
\end{figure}

Figure~\ref{fig:conv} highlights the efficiency of directly computing $X_{\mu\nu}$ and CD from Bloch derivatives, instead of from a sum-over-states approach using momentum matrix elements (Eq.~\ref{eq:XfromP}).
We focus here on convergence, and present comparisons to experiment in the Results section below.

The molecular benchmark, camphor, requires summing more than 1000 bands to converge towards the direct DFT result, which is expected due to the contribution of unbound empty states, while convergence is manageable requiring 100 bands for the solid benchmark, Te. In contrast, direct calculation using $LQ$ matrix elements only requires 50 bands for camphor and 75 for Te. This represents a massive computational speed-up as typical plane-wave DFT calculations scale as $N_b^3$ (limited by subspace diagonalization) when calculating large numbers of empty states, where $N_b$ is the number of electronic bands.
Specifically, for the camphor example, the evaluation of just 50 bands (including $LQ$ matrix elements) is at least 100x faster than computing 1000 bands (without $LQ$ matrix elements) on the same computational resources.
The direct DFT calculation using Bloch derivatives avoids the extra convergence with respect to bands which arises from a sum-over-states approach, and enables prediction of CD regardless of dimensionality and unit cell complexity.

\subsection{Evaluation using Wannier Functions}
\label{sec:WannierLQ}

We use matrix elements of $\vec{p}$, $\vec{S}$ (when using SOC), $\vec{L}$ and $Q_\mu\nu$ computed directly using DFT to predict CD using Eq.~\ref{eq:CD}.
This requires an integration of $\vec{k}$ over the Brillouin zone, which we do using a Monte Carlo sampling, with the matrix elements at each $\vec{k}$ calculated as discussed above.
We choose Monte-Carlo sampling because the $\vec{L}$ and $Q_{\mu\nu}$ matrix elements do not vary smoothly with $\vec{k}$ in a crystalline system, often exhibiting sharp peaks.
Regular $\vec{k}$-meshes may converge a smooth integrand more quickly, but could lead to unknown systematic errors for non-smooth integrands, which would show up instead as statistical errors in Monte Carlo sampling.
Regardless, the technique shown here can be applied easily with any of these Brillouin-zone sampling schemes.

While this is useful to benchmark results systematically, without needing any band convergence, it still requires convergence over the number of $\vec{k}$ used, which eventually becomes computationally expensive.
For example, Figure~\ref{fig:conv}(c) shows that predicting the CD for Te requires over 2000 $k$-points to converge the Brillouin zone integral.

To accelerate calculations further, we also implement a Wannier interpolation alternative to the direct DFT calculation above.
The interpolation of $\vec{p}$ and $\vec{S}$ is straightforward using the standard approach of applying Wannier rotations to the DFT-computed matrix elements, followed by a Fourier transform to a real space version~\cite{Marzari1997}.
The matrix elements can then be reconstructed at arbitrary $\vec{k}$ using an inverse Fourier transform, and then rotated to the eigen-basis using eigenvectors of the Hamiltonian interpolated to the same $\vec{k}$~\cite{Brown2016, Sundararaman2017, Xu2020}.

However, Wannier interpolation of the $X_{\mu\nu}$ needed for $\vec{L}$ and $Q_{\mu\nu}$ requires care because the derivatives in $\partial_\vec{k} u_{\vec{k}n'}$ introduce sharp features in reciprocal space that can lead to interpolation errors.
One possible approach to mitigate this closely mirrors the calculation of the Berry curvature, and combines interpolation of position operators (and higher moments) with computing the derivative of the unitary rotation in Wannier basis~\cite{Yates2007}.
Here, we propose a simpler approach that more closely mimics the direct DFT calculation above.

The key idea is that since $\partial_\vec{k} u_{\vec{k}n'}$ introduces long-range real-space terms after transformation to Wannier functions, we invoke a range separation to only Wannier-interpolate the short-ranged part.
Specifically, we define the `smoothed derivative', 
\begin{equation}
\overline{\partial_\vec{k} u_{\vec{k}n'}} \equiv \sum_{a} 
\partial_\vec{k} \left( \sum_n u_{\vec{k}n} U^\vec{k}{na} \right)
U^{\vec{k}\ast}{n'a},
\end{equation}
where $U^\vec{k}{na}$ are the unitary rotations that maximally localize the Wannier functions~\cite{Marzari1997}.
These rotations match the Bloch functions at nearby $\vec{k}$, thereby eliminating the sharp features in the derivative.
We apply Eq.~\ref{eq:X} using this smoothed derivative instead to get matrix elements $\bar{X}^\vec{k}_{\mu\nu,n'n}$, which can be interpolated readily in exactly the same way as the momentum and spin matrix elements.

After Wannier interpolation to a finer $\vec{k}$ mesh, we must account for the previously eliminated long-range parts, which arise due to the derivative of the Wannier rotations.
Specifically, we get combinations of the form,
\begin{equation}
D^\vec{k}_{\mu,n'n} =
    \sum_{a} U^\vec{k}_{n'a}
    \left( \partial_{k_\mu} U^{\vec{k} \dagger}_{an} \right)
\end{equation}
that corrects the short-ranged $\bar{X}$ to the desired full matrix elements,
\begin{equation}
X^\vec{k}_{\mu\nu} = 
    \bar{X}^\vec{k}_{\mu\nu} + \frac{i}{2} \left(
    D^\vec{k}_\mu p^\vec{k}_\nu + p^\vec{k}_\nu D^\vec{k}_\mu
\right),
\end{equation}
written in terms of matrices in band space (omitting band indices) for simplicity.
We compute the rotation derivative $D^\vec{k}_\mu$ after Wannier interpolation exactly as discussed in detail in~\cite{Yates2007}, where it is applied for Berry curvature and other band derivative calculations.
The main modification here is to simplify the reconstruction process by interpolating a simplified version of $X^\vec{k}_{\mu\nu}$.
Additionally, by using momentum matrix elements computed and interpolated directly (without using $D^\vec{k}_\mu$ as done in~\cite{Yates2007}), we have fewer powers of $D^\vec{k}_\mu$ in our final expression, making this approach less sensitive to convergence of the Wannier transformation with respect to $k$-mesh.
Finally, once we have $X^\vec{k}_{\mu\nu}$ with fine $k$-sampling using the above approach, we can use it to extract $\vec{L}$ and $Q_{\mu\nu}$ and compute CD using Eq.~\ref{eq:CD} just as before.
Figure~\ref{fig:conv}(b) demonstrates the near equivalence of computing CD directly from DFT matrix elements and using Wannier interpolation of the matrix elements for the Te crystal benchmark. Minor discrepancies between the two curves are a result of the difficulties in Wannier interpolation of $\vec{L}$ and $Q_{\mu\nu}$ matrix elements, as mentioned above.

\subsection{Circular dichroism units}

Finally, we discuss units used in reporting CD and point out convenient units and conversions for comparing predicted and measured CD in chiral crystals.
Experimental CD is typically reported as ellipticity introduced into linearly-polarized light because of the differential absorption of the two initially-equal circular-polarized components.
Specifically, after path length $l$ through the sample, the fractional difference in intensity of the two circular components becomes $l\Delta\alpha$, where $\Delta\alpha$ is the Naperian absorbance difference computed by Eq.~\ref{eq:CD}.
This ellipticity corresponds to a phase difference $\theta = l\Delta\alpha/4$ (in radians) between the two linear polarized components.

Experimentally, absorbance is typically defined with a base-10 logarithm, which we'll denote $\Delta\alpha_{10}$, and the angle is reported in degrees, leading to the conversion $\theta/\mathrm{deg} = (180 \ln 10/(4\pi)) \cdot l\Delta\alpha_{10} \approx 32.982 \cdot l\Delta\alpha_{10}$ that is commonly used~\cite{Purdie1989}.
Note that the corresponding conversion for the Naperian absorbance is $\theta/\mathrm{deg} = (45/\pi) \cdot l\Delta\alpha \approx 14.324 \cdot l\Delta\alpha$.

For molecules, CD is typically reported as molar circular dichroism, $[\theta] = \theta/(Cl)$, where $C$ is the molar concentration of the chiral molecule.
In terms of typical experimental units used,
\begin{equation}
\frac{[\theta]}{\mathrm{deg}\cdot\mathrm{cm}^2/\mathrm{dmol}} 
= \underbrace{\frac{4500 \ln 10}{\pi}}_{\approx 3298.2}
    \frac{\Delta\alpha_{10} / (\mathrm{cm}^{-1})}{C / (\mathrm{mol/l})}.
\end{equation}
For our theoretical predictions, the concentration is one molecule in a unit cell of volume $\Omega$, yielding
\begin{equation}
\frac{[\theta]}{\mathrm{deg}\cdot\mathrm{cm}^2/\mathrm{dmol}} 
= \underbrace{\frac{4500 N_A/\mathrm{mol}^{-1}}{10^{27}\pi}}_{\approx 0.86261}
    \frac{\Delta\alpha}{(\mathrm{cm}^{-1})} \cdot \frac{\Omega}{\AA^3},
\end{equation}
where $N_A$ is the Avogadro number.
We convert our predictions using Eq.~\ref{eq:CD} to these units for molecules, as the units are well-standardized.

For crystals, CD may be reported as ellipticity per path length through the crystal,
\begin{equation}
\frac{\theta/l}{\mathrm{deg/mm}} = \underbrace{\frac{45}{10\pi}}_{\approx 1.4324} \frac{\Delta\alpha}{\mathrm{cm}^{-1}},
\label{eq:CrystalCDunits}
\end{equation}
or converted to the Naperian absorbance difference as shown on the right side of the above equation.
Either of these choices are appropriate, as they can easily be converted as shown above.
However, several experimental reports of CD in crystals report the ellipticity without reporting the path length $l$, making it impossible to compare quantitatively to theoretical predictions.
Below, we report results for crystals in deg/mm for direct quantitative comparison with experiment.
When comparing to experiments that report ellipticity without path length, we report the path length that would lead to best match with our theoretical predictions, so that we retain the absolute scale on our predictions.
 \section{Methods}

We implemented the DFT and Wannier calculation of required matrix elements in the open-source plane-wave DFT software JDFTx~\cite{Sundararaman2017}.
We use scalar and full-relativistic norm-conserving pseudopotentials respectively for calculations with and without spin-orbit coupling from PseudoDojo~\cite{vanSetten2018}, with a kinetic energy cutoff of 45~$E_h$ for the plane wave basis.
We use the Perdew-Burke-Ernzerhof (PBE) exchange-correlation functional for all crystalline materials~\cite{perdew1996generalized}, and the Becke 3-parameter Lee-Yang-Parr (B3LYP) hybrid exchange-correlation functional for the molecules~\cite{Tirado2008}.
We use initial structures from the Materials Project database~\cite{Jain2013} unless indicated otherwise, and fully relax lattice vectors (for crystals) and atomic positions for all calculations.
For the hybrid perovskite crystal, we include DFT-D2 dispersion corrections~\cite{Grimme2006} during structural relaxation to account for long-range interactions between the organic components. 
See supplemental information for the converged structures and additional computational details~\cite{SI}.

For Brillouin zone integration in the crystal calculations, we use a $\Gamma$-centered \mesh{12}{12}{12} $k$-mesh in the initial self-consistent calculations, except for the hybrid perovskite, where we use \mesh{8}{8}{8} on account of the significantly larger unit cell.
For direct DFT calculations of the matrix elements, we perform non-self-consistent calculations for varying number of randomly-sampled $k$ in the Brillouin zone (reported with the results).
For Wannier-interpolated calculations, we construct Wannier functions to cover an energy window ranging from the bottom of the set of valence bands to at least 10~eV above the Fermi level / conduction band minimum for metals / semiconductors, starting with random Gaussian orbitals to generate the initial guess for the Wannier rotations~\cite{Marzari1997}.
We use these Wannier functions to interpolate matrix elements to a fine Monte-Carlo sampling of $k$ within the Brillouin zone, with the number of $k$ for each system reported along with the results.
 \section{Results}

\subsection{Molecular benchmarks}

\begin{figure}
\includegraphics[width=\columnwidth]{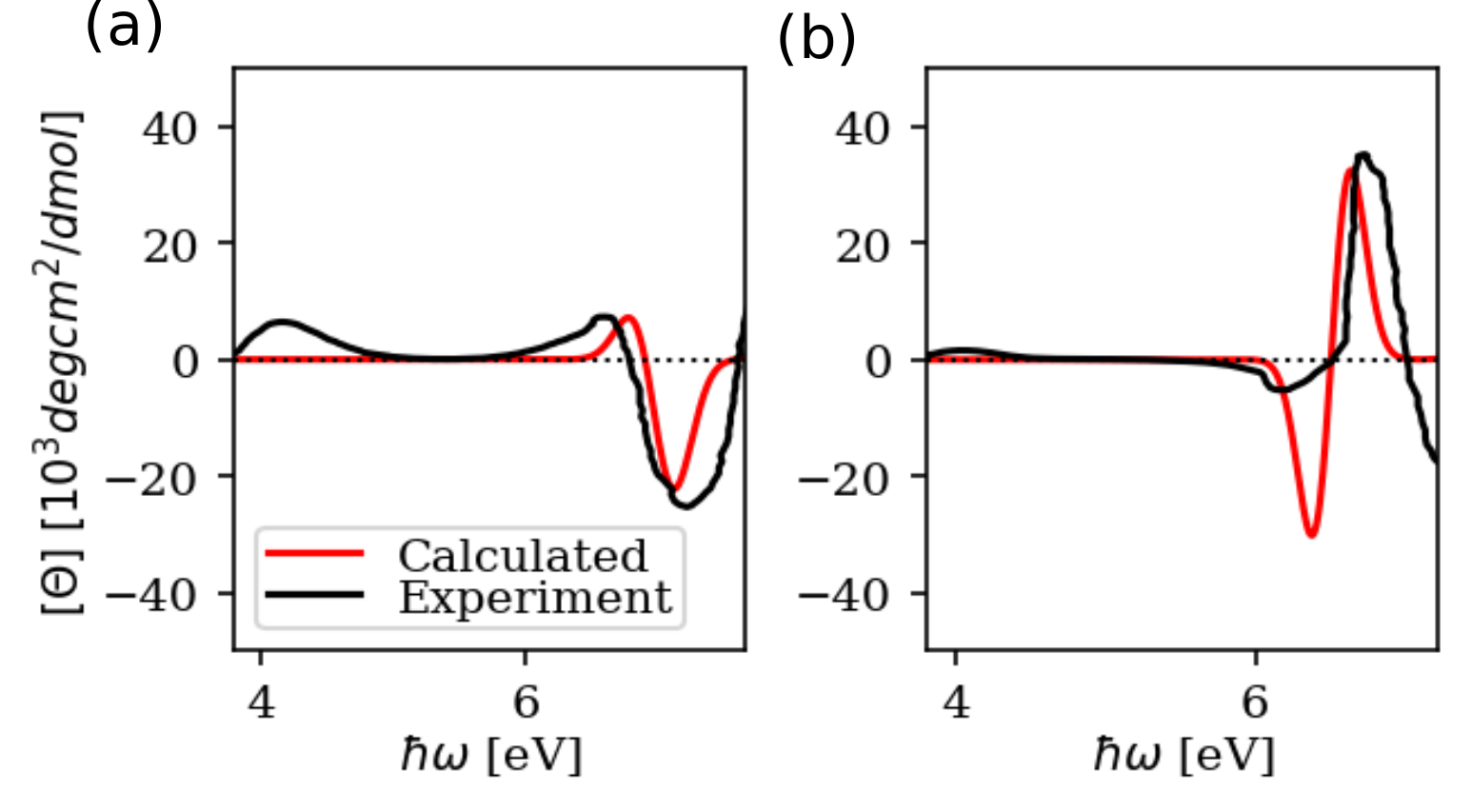}
\caption{Calculated CD spectra for (a) camphor and (b) norcamphor agree well with measurements~\cite{Diedrich2003}.
The HOMO-LUMO gap is corrected using the scissor operator (shift unoccupied energy levels) to match experiment, but the CD spectrum is matched quantitatively without any scaling.}
\label{fig:mol}
\end{figure}

We begin by benchmarking our approach for prototypical chiral molecules, given the abundance of experimental CD data.
Figure~\ref{fig:mol} shows reasonable agreement of predicted CD with experimental measurements for camphor and norcamphor~\cite{Diedrich2003}.
We use the B3LYP hybrid exchange-correlation functional for these molecules, as utilized in previous molecular CD predictions~\cite{Goerigk2009}.
For these predictions, we use the direct DFT calculation of $\vec{L}$ also outlined in Section~\ref{sec:DirectLQ}, as there is no need for Brillouin zone sampling and Wannier interpolation for molecules.
Note that the quadrupole $Q_{\mu\nu}$ does not contribute to the trace of Eq~\ref{eq:CD}.
We apply a scissor shift of the unoccupied states of 1.1 and 0.6~eV respectively to match the HOMO-LUMO gap to experiment, and apply a Gaussian broadening of 0.15~eV to the discrete transitions, but the magnitude of CD is compared directly with no arbitrary scaling.
Note the strong agreement of both the signs and magnitudes of peaks for both molecules. 
However, the first peak around 4 eV in the experimental spectra is missing in our calculations, while previous TD-DFT calculations capture this peak~\cite{Diedrich2003}.
This likely arises from our method being based on an independent-particle approximation using DFT, while TD-DFT includes local-field effects (Hartree term) in the excitation.

\subsection{Isotropic CD from cubic crystals}

\begin{figure}
\includegraphics[width=\columnwidth]{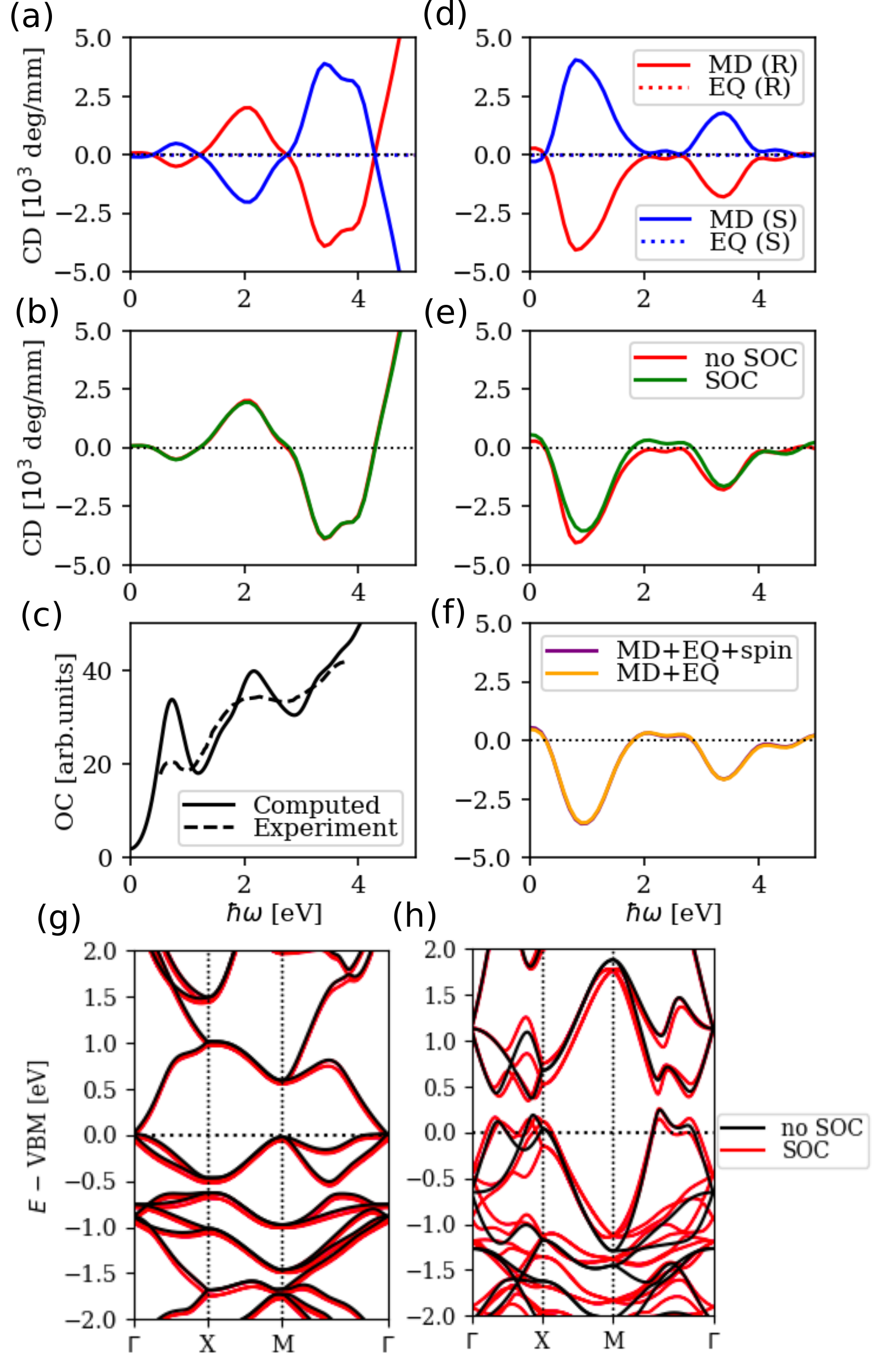}
\caption{(a) Predicted CD for cubic chiral semimetal CoSi is entirely from the magnetic dipole (MD) terms, with no contribution from the electric quadrupole (EQ) terms.
The CD is equal and opposite for enantiomers with the P2\sub{1}3 (R) and P2\sub{1}$\bar{3}$ (S) space groups, and (b) is unaffected by spin-orbit coupling (SOC).
(c) Computed optical conductivity (OC) for CoSi shows good agreement with experiment in the location of the absorption peaks, and  the differential absorption for CD also peaks at the same frequencies.
(d,e) Same as (a,b) for MgPt, showing a non-negligible impact of SOC.
(f) Spin contributions do not change the CD of MgPt, and the impact of SOC is instead due to band energy changes.
(g) Band structure of CoSi is largely unchanged by SOC, while (h) that of MgPt shows spin-splits $\sim 0.1 - 0.2$~eV.
}
\label{fig:iso}
\end{figure}

Next, we consider crystalline systems with symmetries that lead to isotropic CD, where the quadrupole contribution remains zero.
Specifically, we investigate cubic crystals CoSi and MgPt, with the P2\sub{1}3 chiral space group (and enantiomer P2\sub{1}$\bar{3}$) that been the focus of recent research in spin properties~\cite{Tan2022}.
CoSi is a chiral semi-metal with promise for topologically-protected surface conduction for integrated circuit technologies~\cite{Chen2020}.
Figure~\ref{fig:iso}(a) shows that CoSi exhibits CD starting at zero photon energy, as expected since it is a semimetal.

The CD intensities peak near 0.75 and 2.0~eV, coincident with the corresponding peaks in the computed optical conductivity (FIG.~\ref{fig:iso}(c)), which in turn agrees well with the experimental absorption spectrum~\cite{Marel1998, Kudryavtsev2007}.
This agreement between peak frequencies of overall absorption in OC and differential absorption in CD arises from a peak in joint density of states for absorption, but not all OC peaks need to be CD peaks in general due to the differences in relevant matrix elements ($\vec{p}$ alone vs. also accounting for $\vec{L}$ and $Q_{\mu\nu}$).

CD measurements have not yet been reported for CoSi, and would be a useful test for chiroptical properties in topological materials in future work.
Predictions of CD with and without SOC differ negligibly for CoSi (FIG.~\ref{fig:iso}(b)), as expected since it is composed entirely of light atoms 
and its band structure changes negligibly with inclusion of SOC (FIG.~\ref{fig:iso}(g)).

To elucidate the effect of spin-orbit coupling, we also predict CD for MgPt, incorporating heavy element Pt in the same chiral crystal structure as CoSi (FIG.~\ref{fig:iso}(c)).
MgPt also exhibits CD for photon energies starting from zero, and of a similar magnitude to CoSi (FIG.~\ref{fig:iso}(d)).
The CD predictions with and without SOC are qualitatively similar, but with a non-negligible change in magnitudes and peak positions (FIG.~\ref{fig:iso}(e)).
To elucidate how SOC changes the CD, Figure~\ref{fig:iso}(f) shows the CD predicted with and without the spin contribution from the terms containing $\vec{S}^\vec{k}_{n'n}$ in Eq.~\ref{eq:CD}.
These CD spectra are indistinguishable, indicating that the spin contribution is negligible even in this high-SOC material.
Instead, the small effect of SOC is primarily due to changes in band energies

by $0.1 - 0.2$~eV (FIG.~\ref{fig:iso}(h)), leading to small shifts in the CD peaks and magnitudes.
Since the CD changes negligibly even in systems with sizable SOC, we omit SOC for most of the test systems considered next.

\subsection{Anisotropic CD from hexagonal crystals}

\begin{figure}
\includegraphics[width=\columnwidth]{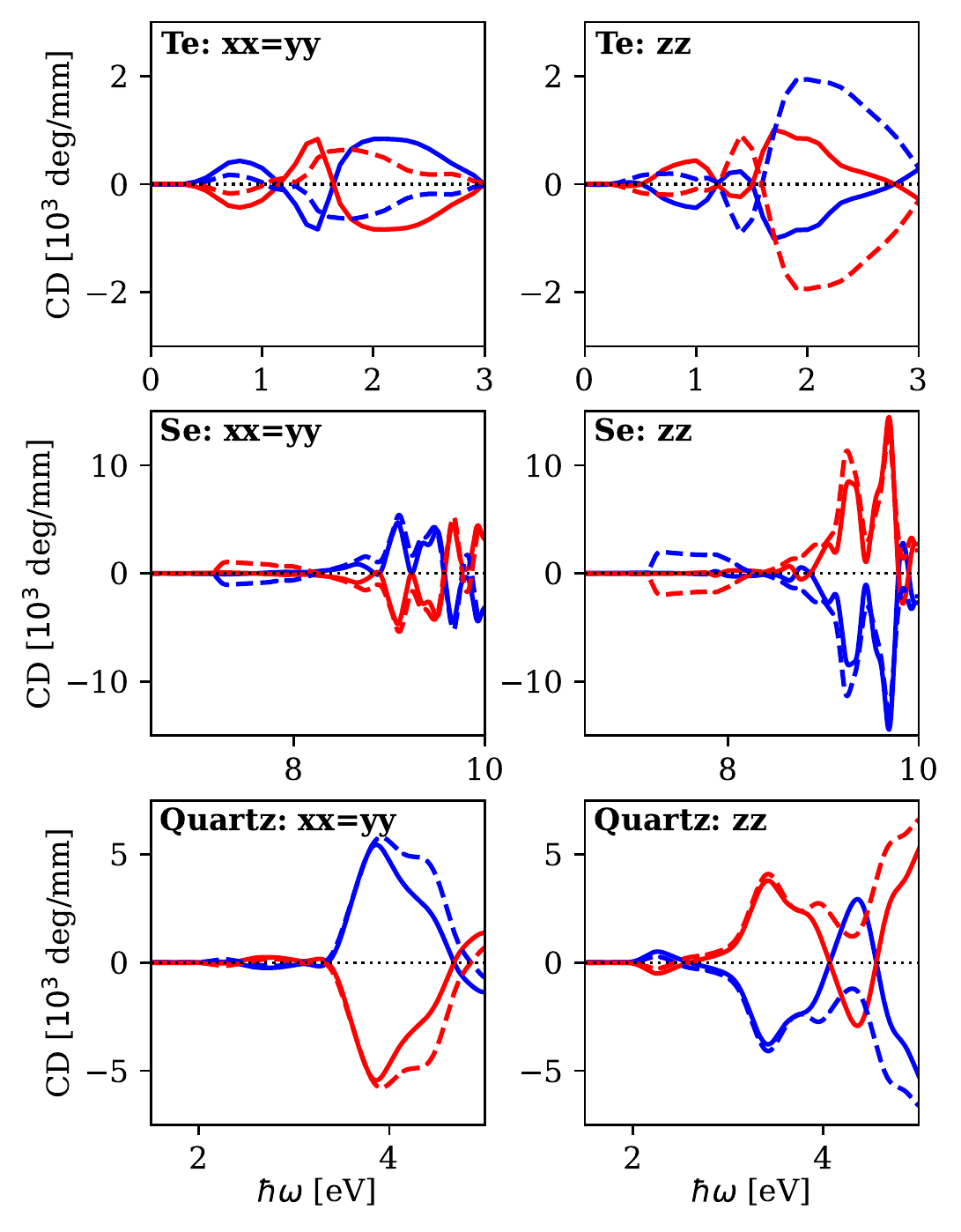}
\caption{Calculated CD tensor of hexagonal crystals quartz, Te, and Se shows strong anisotropy between in-plane (xx=yy) and out-of-plane  (zz) directions.
The total CD (MD+EQ) is substantially modified by the electric quadrupole (EQ) contributions compared to the magnetic-dipole-only CD (MD).}
\label{fig:aniso}
\end{figure}

To study the impact of the electric quadrupole contributions and anisotropic CD, we next consider hexagonal crystals with two distinct values in the CD tensor: $xx = yy$ for in-plane wave propagation, distinct from the $zz$ component for out-of-plane propagation and all off-diagonal terms zero.
Specifically, we calculate quartz (SiO\sub{2}), trigonal Se and trigonal Te, all of which are in the P3\sub{1}21 and P3\sub{2}21 space groups respectively for the left- and right-handed enantiomers.
For Te, we include spin-orbit coupling because this introduces a small gap (0.03~eV), which we can correct with a scissor to the experimental gap of 0.34~eV, while the no-SOC calculation is metallic and cannot be corrected~\cite{Wang2023}.
Figure~\ref{fig:aniso} shows that the electric quadrupole (EQ) contributions substantially alter the total predicted CD for all systems.
In most cases (except $xx=yy$ component for Se), there is a partial cancellation between the magnetic dipole (MD) and EQ contributions, leading to the total CD being of smaller magnitude than predicted using the MD component alone.
This showcases the necessity to include the quadrupole contributions for correct general treatment of CD in crystals.

To compare with experimental measurements, it is preferable to directly compare CD, as this provides clear energy-dependent signatures that often change in sign, as seen for the molecules above.
However, in the absence of CD measurements, we can also compare to optical rotation (OR).
OR arises from phase difference on the propagation of LC and RC light due to differences in $\Re\epsilon(\omega)$ below the band gap, in contrast to differential absorption from $\Im\epsilon(\omega)$ that leads to CD.
Consequently, the Kramers-Kronig relation yields the OR~\cite{Krykunov2006},
\begin{equation}
\theta(\omega) = \frac{2\omega^2}{\pi} \int_{0}^{\infty} d\omega'
    \frac{\Delta\alpha(\omega')}{\omega'(\omega'^2-\omega^2)},
\end{equation}
from the CD, $\Delta\alpha$, computed using Eq.~\ref{eq:CD}.
We can covert OR to experimental units (deg/mm) using Eq.~\ref{eq:CrystalCDunits} in the same way as for CD.

Quartz has no reported CD measurements, due to the high band gap $\sim 9$~eV that would necessitate deep ultraviolet measurements with wavelength $< 140$~nm.
Additionally, while optical rotation measurements for quartz have been reported~\cite{Gutowsky1951}, they do not provide a reliable test for the CD spectrum since they are too far below the band gap and relatively weak.

\begin{figure}
\includegraphics[width=\columnwidth]{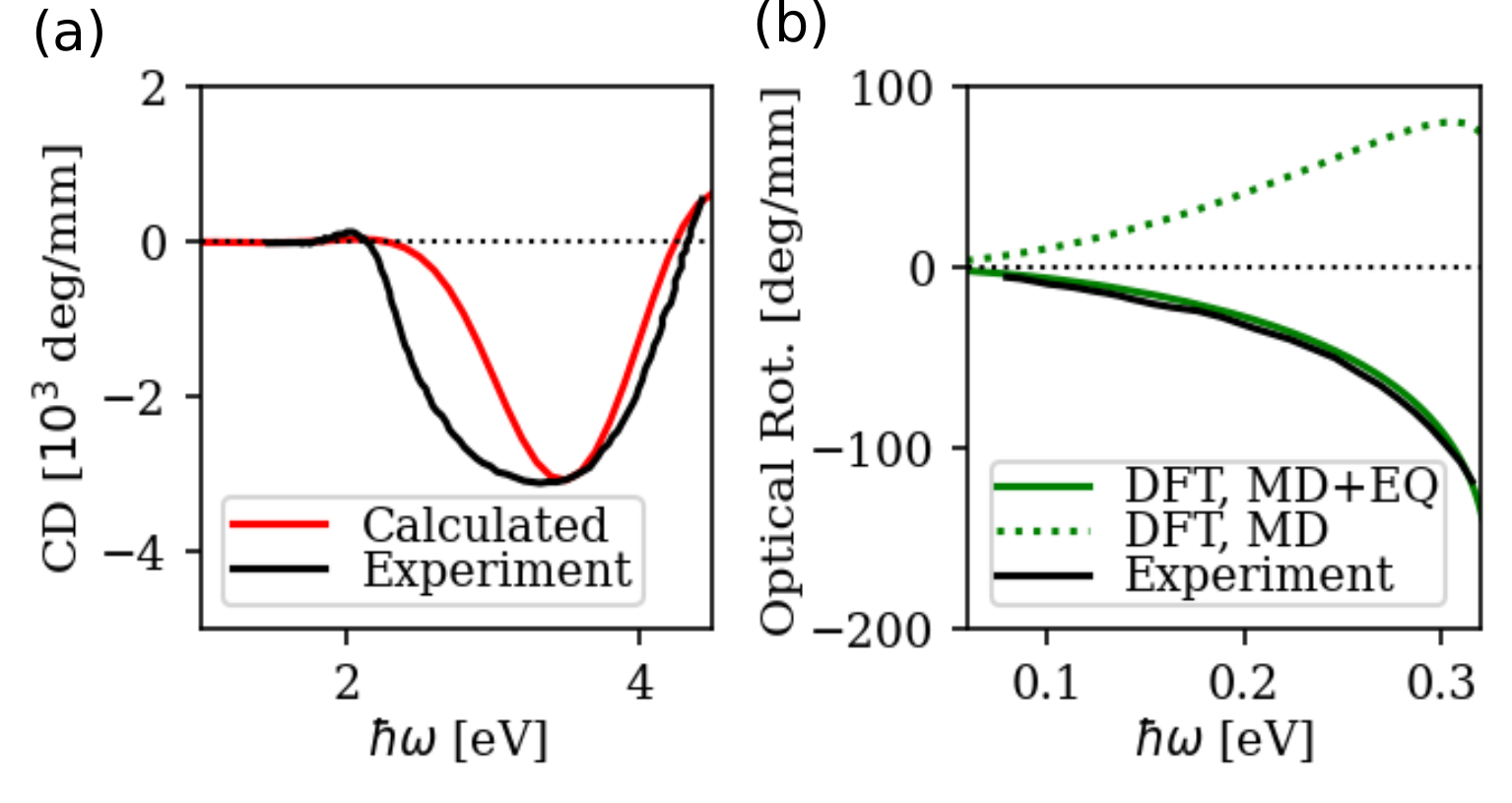}
\caption{(a) Predicted CD for Se matches experiment well~\cite{Saeva1977}, after using a scissor to correct the band gap, assuming path length $l=0.01$~mm not reported in~\cite{Saeva1977}.
(b) Optical rotation (OR) for Te, obtained by Kramers-Kronig integration of predicted CD, matches experimental measurements quantitatively~\cite{Ades1975}.
The electric quadrupole term is critical to even get the correct sign as experiment for the OR of Te.}
\label{fig:SeTe}
\end{figure}

Trigonal selenium has served as a prototypical elemental semiconductor for investigating chirality~\cite{Pal2013, Caldwell1958}.
Figure~\ref{fig:SeTe}(a) shows that our CD predictions for Se are in good qualitatively agreement with experimental measurements~\cite{Saeva1977}, in the sign as well as the relative magnitude of the peaks.
Here, we have used a scissor of 0.97~eV to correct the computed band gap to the experimental value of 2.0~eV.
Additionally, the experimental CD reported the overall ellipticity, without crystal dimensions or path length, leading to the overall magnitude not being directly comparable.
We assume a path length $l=0.01$~mm in the experiment, which brings the experimental CD to the correct scale compared to our predictions.

Trigonal tellurium is one of the most studied chiral crystals, showing promise for chiral optoelectronic properties such as the circular photogalvanic effect~\cite{Tsirkin2018}.
Experimental CD for Te has so far been restricted to polycrystalline samples with various degrees of chiral purity, leading to large sample-to-sample variations in the location, magnitude, and sign of CD peaks~\cite{BenMoshe2014}.
Consequently, Figure~\ref{fig:SeTe}(b) compares to optical rotation measurements along the $c$-axis ($zz$ tensor component) of Te single crystals~\cite{Ades1975}, and finds quantitative agreement.
Here, we have a scissor of 0.31~eV to correct the gap, and the magnitude of the OR is then compared absolutely with no arbitrary scaling.
Note that the quadrupole contribution is essential here even to get the sign of the OR correct.
This can be seen earlier in Figure~\ref{fig:aniso} as well, where the lowest energy total CD for Te has opposite signs between in-plane and out-of-plane propagation, while the MD-only CD having the same sign, showing that this extreme anisotropy is introduced by the quadrupole (EQ) component.

\subsection{CD in complex chiral crystals}

\begin{figure}
\includegraphics[width=\columnwidth]{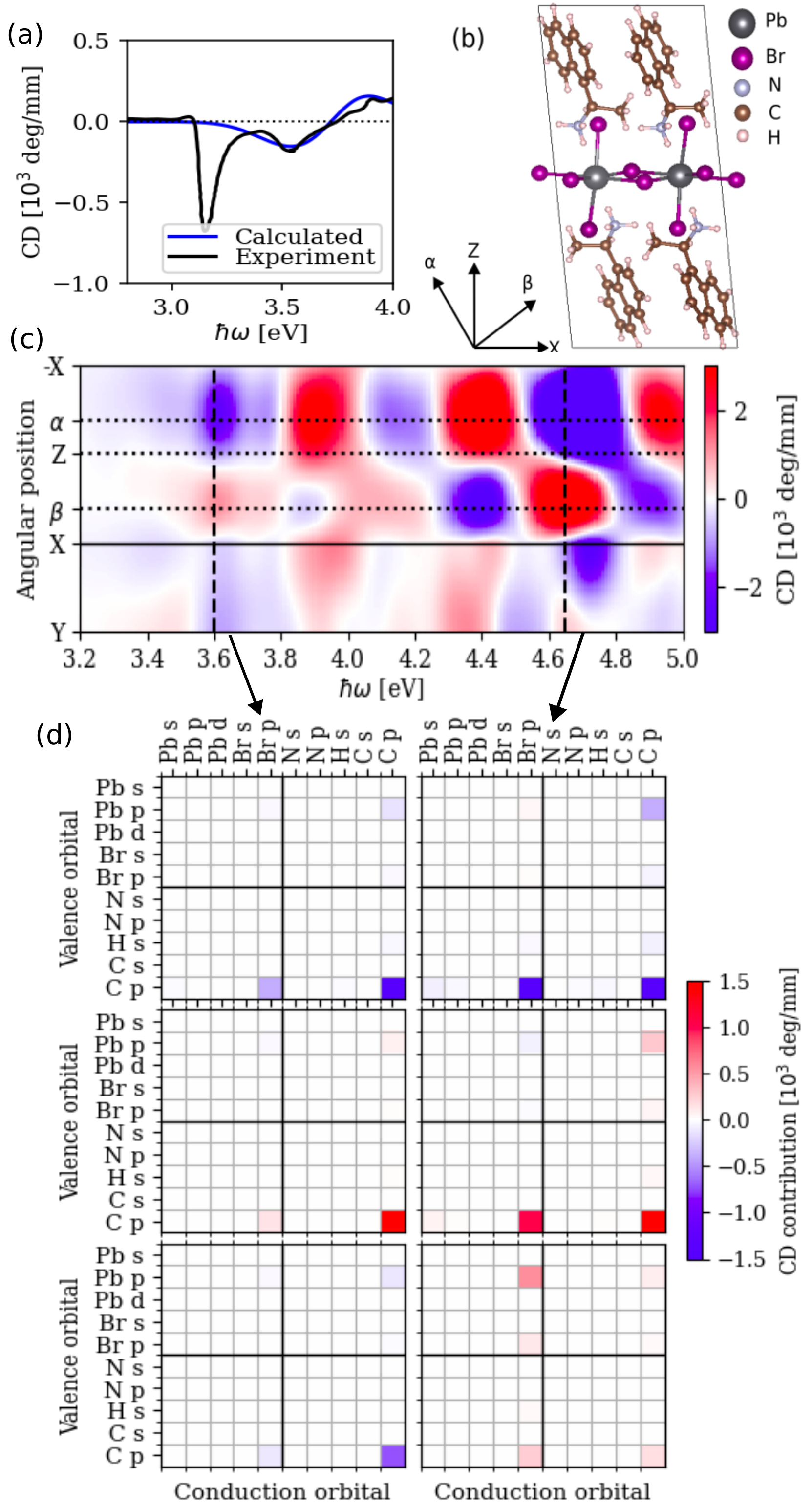}
\caption{(a) Predicted CD agrees with measurements~\cite{Jana2020} for hybrid perovskite (S)-NPB (1-(1-naphthyl)ethylammonium lead bromide) beyond the first excitonic peak not captured in our DFT calculations.
Since experimental data is polycrystalline and does not report path length, we show the isotropic average (Tr$\Delta\alpha$/3) and fit path length $l=0.02$~mm to match the predicted magnitude.
(b) Atomic structure of (S)-NPB, where the chiral molecules distort the inorganic lattice to induce chirality throughout, indicating eigendirections $\alpha$ and $\beta$ in the $xz$-plane that lead to maximal CD magnitude.
(c) Angular dependence of CD in the $xz$- and $xy$-planes, showing maximum magnitudes along the $\alpha$ and $\beta$ directions for all frequencies, almost an an order of magnitude higher than that along the optical axis ($y$ direction).
(d) Orbital decomposition of CD along the eigendirections at two peak frequencies $A$ and $B$: each heat map is split into quadrants based on transitions from initial and final states in the organic molecule and inorganic lattice components.
The intramolecular contributions dominate the CD, with mixed and inorganic contributions becoming important only for higher frequencies in the $y$ direction.}
\label{fig:expNPB}
\end{figure}

Finally, to showcase the capability of first-principles CD calculations of complex materials, we consider chiral hybrid perovskites.
Hybrid organic-inorganic perovskites provide an interesting platform to control crystal symmetry using molecular structure to target exciting optical properties~\cite{Dang2020, Ahn2020}.
In particular, introducing chiral molecules into hybrid perovskites transfers the chirality to the crystal structure overall, distorts the inorganic lattice and enhances the chiral response overall~\cite{Hu2020, Jana2020}.

CD measurements have been recently reported for several chiral hybrid perovskites, typically with complex structures and large unit cells with 100 - 200 atoms per primitive cell.
As an example here, we focus on 1-(1-naphthyl)ethylammonium lead bromide (NPB) with 115 atoms in the primitive cell, with structure and CD characterized extensively in Ref.~\cite{Jana2020}.
Figure~\ref{fig:expNPB} compares our predicted CD with experimental measurements from Ref.~\cite{Jana2020}.
We use a scissor of 0.6~eV to rigidly shift the conduction band energies relative to the valence band energies in order to match the band gap corresponding to the experimental absorption edge.
Additionally, the experiment reported ellipticity without path length, so we fit the path length $l = 0.02$~mm for best match of CD magnitude with our predictions.
We find good agreement with the experimental CD, except for missing the excitonic peak at $\hbar\omega = 3.2$~eV.
Such hybrid perovskites are well-known to exhibit strong excitonic contributions to the absorption~\cite{Jana2020, Sercel2020}, especially up to approximately 0.5~eV above the absorption edge~\cite{Wang2021}.
With many-body effects not accounted for in our DFT calculations (which are at the mean-field level), this excitonic peak is missed as expected, and our results agree with the experimental CD for higher photon energies.

Leveraging the tensorial nature of CD, we next examine the orientation dependence of CD by diagonalizing the tensor to extract the principal directions.
We find that in addition to the optic axis, the $y$ direction, which is an eigenvector by symmetry, the remaining two eigenvectors in the $xz$-plane, which we label $\alpha$ and $\beta$ in Figure~\ref{fig:expNPB}(b)  are relatively insensitive to frequency over the relevant range and located at $-32^\circ$ and $58^\circ$ relative to the $z$-axis, respectively.
Interestingly, eigendirection $\alpha$ is approximately in the plane of the (S)-NEA molecules, while $\beta$ is perpendicular to it and the optic axis ($\hat{y}$) is aligned with the screw axis of the inorganic lattice. 
We find that the CD along $y$ peaks at approximately the same frequencies as $\alpha$ and $\beta$ (Fig.~\ref{fig:expNPB}(c)), but with a much lower magnitude.

Finally, we decompose the CD predictions by orbital contributions to investigate whether the organic or inorganic components of 2D HOIPs provide a stronger contribution to CD.
To do this, we weight the contribution of each transition ($\vec{k}nn'$ combination) in Eq.~\ref{eq:CD} by atomic orbital projections of the initial and final state wavefunctions.
For $\alpha$ and $\beta$ directions, we find the dominant contribution to CD at all frequencies to be C$_p$-C$_p$ transitions (where C$_p$ denotes carbon $p$ orbitals), with minor contributions from C$_p$-Br$_p$ and Pb$_p$-C$_p$ transitions (Fig.~\ref{fig:expNPB}(d)). 
The relative strength of these minor contributions tends to increase with frequency, as is demonstrated in the CD peaks located at 3.6eV and 4.65eV, labeled $A$ and $B$, respectively.
The $y$ direction shows the same relative contributions for the $A$ peak as in the $\alpha$ and $\beta$ directions, but for the $B$ peak, is dominated by Pb$_p$-Br$_p$ transitions instead.
This dominant inorganic contribution makes sense since the $y$ direction corresponds to the screw axis of the inorganic component of the perovskite.
Overall, the chiral molecules dominate the CD in 2D HOIPs, while transitions involving the inorganic lattice contribute at specific frequencies and directions.

 \section{Conclusions}

We have developed a unified computational framework for first-principles prediction of circular dichroism in both molecular and periodic systems, motivated by the increasing recent interest in chiral crystals for their unique spin and optoelectronic properties.
We find excellent agreement between our predictions and experimental measurements for both molecular benchmarks and crystals ranging in complexity from elemental semiconductors to hybrid organic-inorganic perovskites.
We demonstrate the versatility of our method by testing it for systems with varying degrees of anisotropy and unit cell complexity.
Critically, this versatility is enabled by direct DFT and Wannier evaluation of the orbital angular momentum and electric quadrupole matrix elements without requiring a sum over all states, substantially reducing computational cost and convergence challenges for large unit cells and low-dimensional systems.

Further, developing CD as a rank-2 tensor with respect to propagation direction facilitates intuitive analysis of the anisotropic CD from complex crystals.
We show that the quadrupole contribution substantially modifies the anisotropic CD for all the non-cubic crystals we studied, and is essential for quantitative agreement with single-crystal CD measurements.
We find that SOC affects CD primarily by modifying the electronic structure and band energies, rather than by a direct contribution from the spin matrix elements.
Leveraging the capability to compute CD tensors for complex materials, we perform a thorough analysis of the directional dependence of CD using a 2D hybrid perovskite as an example, identifying the unit cell directions along which CD is maximized.
Further, decomposing the CD signal into organic, inorganic, and mixed contributions allows us to conclude that the chiral molecules are dominate the optical activity in the relevant frequency range (excluding excitonic transitions), whereas the inorganic lattice dominates only in certain directions and at higher frequencies.

Finally, we discuss a standardization necessary in experimental reports of CD in crystals.
We recommend the use of either differential Naperian absorbance in inverse length units or ellipticity change per unit path length in units such as deg/mm.
Absolute ellipticity (\textit{e.g.}, in degrees) reported without crystal dimensions or path length, which has been prevalent in recent experimental work, precludes quantitative comparison and should be avoided.

Specifically, we recommend reporting path length with at least 10\% accuracy, based on the precision now achievable for first-principles CD predictions.

This is critical as an increasing number of complex chiral crystals such as hybrid perovskites are being developed experimentally, and can now be predicted with the high-throughput-capable first-principles approach presented here.\\
 
\section*{Acknowledgements}
This work was supported by the Department of Energy, Basic Energy Sciences, under grant \#DE-SC0023301.
Calculations were carried out at the Center for Computational Innovations at Rensselaer Polytechnic Institute,  and at the National Energy Research Scientific Computing Center (NERSC), a U.S. Department of Energy Office of Science User Facility located at Lawrence Berkeley National Laboratory, operated under Contract No. DE-AC02-05CH11231 using NERSC award ERCAP0023917.

\bibliographystyle{apsrev4-2}
 
\end{document}